\documentclass[aps,prb,superscriptaddress,showpacs,10pt,twocolumn]{revtex4-2}

\usepackage[T1]{fontenc} 
\usepackage[utf8]{inputenc} 
\usepackage{lmodern} 
\usepackage{microtype} 
\usepackage{bm}
\usepackage{amssymb}
\usepackage{graphicx}
\usepackage{multirow}
\usepackage{colortbl}
\usepackage{mathtools}

\newcommand{\ieap}{Institut f\"ur Experimentelle und Angewandte Physik, Christian-Albrechts-Universit\"at zu Kiel, D-24098 Kiel, Germany}
 
\newcommand{\didv}{$dI/dV$}
\newcommand{\up}{PbPc$\uparrow$}
\newcommand{\dn}{PbPc$\downarrow$}
\newcommand{\hnull}{H$_0$Pc}
\newcommand{\hzwei}{H$_2$Pc}

\newcommand{\ie}{\emph{i.\,e.}}
\newcommand{\eg}{\emph{e.\,g.}}

\begin{document}

\title{Making Closed-Shell Lead-Pthalocyanine Paramagnetic on Pb(100)}

\author{Jan Homberg} \affiliation{\ieap}
\author{Alexander Weismann} \email{weismann@physik.uni-kiel.de}\affiliation{\ieap} 
\author{Richard Berndt}  \email{berndt@physik.uni-kiel.de}\affiliation{\ieap}

\begin{abstract}
Lead phthalocyanine (PbPc), a non-planar molecule, is studied on Pb(100) using scanning tunneling spectroscopy. 
A rigid shift of the molecular orbitals is found between molecules with the central Pb ion pointing to (\dn) or away (\up) from the substrate and understood from the interaction between the molecules and their image charges.
Inside the superconducting energy gap, Yu-Shiba-Rusinov (YSR) resonances are observed for \up\ molecules in islands indicating the presence of a magnetic moment. 
Such bound states are neither present on \dn\ molecules nor isolated \up\ or molecules that lost the Pb ion during deposition (\hnull).
The YSR energies vary depending on the orientation and type of the molecular neighbors.
We analyze the role of the out-of-plane dipole moment of PbPc.
\end{abstract}

\maketitle

\section{Introduction}

Molecular spintronic devices at the limit of miniaturization involve magnetic complexes in contact with conducting electrodes \cite{Meded2011}.
The close proximity of a substrate, however, may stint molecular functions through electronic and steric interactions and can lead to fragmentation \cite{Gruber2020, Yazdani2023}.
On the other hand, the effect of a substrate can be more favorable.
For example, different spin states resulting from charge transfer have been reported from metal-free prophyrin on Au(111) \cite{Zhao2020} and from aza-triangulene molecules on the (111) surfaces of Ag and Au \cite{Wang2022}.
Spin-state switching of closed-shell retinoic acid was observed on Au(111) and interpreted in terms of a sigmatropic reaction \cite{Karan2016, Bocquet2018}.
The importance of molecule-substrate interactions for spin-crossover effects was repeatedly highlighted \cite{Gopakumar2012, Miyamachi2012, Malavolti2018, Fourmental2019, Liu2020, Johannsen2021}.
The diamagnetic closed-shell molecule phthalocyanine (\hzwei) and aluminum phthalocyanine (AlPc) were found to become paramagnetic on a Pb substrate \cite{Homberg2020, Li2022}.
This striking effect has been attributed to electrostatic interactions in molecular arrays and was not observed from isolated adsorbed molecules.
Tuning of the intermolecular interactions may therefore be used to control molecular magnetism in the latter cases.

The tool used to probe the magnetism of single phthalocyanine molecules is scanning tunneling spectroscopy of Yu-Shiba-Rusinov (YSR) resonances \cite{yu_bound_1965, shiba_classical_1968, rusinov_theory_1969}.
These resonances arise from the interaction of a magnetic impurity with the Cooper pairs of a superconductor and appear as prominent pairs of peaks at energies $\pm E_\mathrm{YSR}$ in the excitation gap of the superconductor.
This effect has been investigated from magnetic atoms and molecules \cite{yazdani_probing_1997, franke_competition_2011, hatter_scaling_2017, heinrich_single_2018, brand_electron_2018, kezilebieke_coupled_2018, malavolti_tunable_2018, kezilebieke_observation_2019, Lu2021, RubioVerdu2021, Shahed2022,Vaxevani2022,Homberg2022,Homberg2023,Trivini2023}, partially as a platform for constructing systems that support Majorana modes \cite{NadjPerge2013,Jeon2017,Kim2018,Steiner2022}.

So far, \hzwei\ and AlPc have been discovered to acquire a spin under the influence of suitably arranged nearest neighbors.
Both molecules are planar and exhibit charge imbalances in the molecular plane. 
Here we report on non-planar lead pthalocyanine (PbPc).
This complex features a vertical dipole moment, which is known to have implications for the electronic states of the system \cite{yw10, Gerlach2011, Yuan2013, Huang2013}.
Similar to tin pthalocyanine (SnPc) on Ag(111) \cite{Lackinger2002, Stadler2006, Woolley2007, Wang2009b, Wang2009a}, PbPc may adsorb on Pb(100) with the central ion pointing to (\dn) or away (\up) from the substrate.
We observe a rigid shift of the molecular orbitals between these orientations.
We also find YSR resonances, but only for \up\ molecules that are embedded in dense arrays. 
Variations of the YSR energies with the orientations and types of the molecular neighbors are experimentally explored.

\section{Experimental Methods}

Experiments were performed with two scanning tunneling microscopes operated in ultrahigh vacuum (UNISOKU USM-1300 and a home-built instrument).
Sample temperatures between 1.6 and 4.2\,K were used.
Pb(100) single crystals were prepared by Ar-ion sputtering and annealing to approximately $230^\circ$C\@. 
STM-tips were cut from a Pb wire and sputtered in ultrahigh vacuum.
Submonolayer coverages of PbPc molecules were deposited from a Knudsen cell onto the Pb surface, which was close to ambient temperature.
	
\section{Geometric Structure}

\paragraph{Observed Types of Molecules}

\begin{figure}
\centering \includegraphics[width=0.5\textwidth]{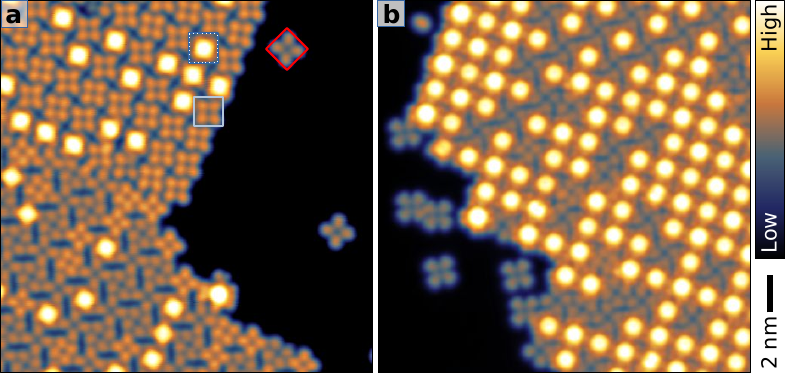}
\caption{: Topographs of submonolayer coverages of PbPc on Pb(100).
(a) and (b) show images from two separate experimental runs.
The PbPc molecules assemble into ordered islands.
In addition, isolated \hnull\ molecules are found.
Squares in dark blue with white dots, cyan, and red indicate examples of \up, \dn, and \hnull\ molecules, respectively. 
In the upper left and lower left parts of (a), two molecular domains are discernible.
Imaging parameters: (a) $I = 200$\,pA, $V = -10$\,mV (b) $I = 100$\,pA, $V = 100$\,mV\@.}
\label{topo}
\end{figure}

After the deposition of submonolayer amounts of PbPc we found molecules in a number of structures, namely isolated molecules, regular arrays of molecule-covered steps and molecular islands.
We observed mainly three types of molecules (Fig.~\ref{topo}).
Intact molecules are adsorbed with their Pb ion pointing to the vacuum side (\up) or to the substrate (\dn) as previously observed from non-planar SnPc \cite{Wang2009b}.
Moreover, some molecules lost the central Pb ion and are denoted \hnull\ below.

While \up\ molecules are easily recognized by their protruding Pb atom, the differences of \dn\ and \hnull\ molecules are rather subtle with \dn\ molecules appearing slightly higher than \hnull.
For identification of \hnull, we used manipulations with the STM tip.
At elevated currents, \up\ may be demetallized like SnPc \cite{Sperl2011} leaving \hnull\ behind.
Alternatively, we prepared \hnull\ by removing the central H atoms from \hzwei\ \cite{meta_2011}.
In both cases, images of the remaining \hnull\ were identical to those of the molecules identified as \hnull\ after PbPc deposition.
Furthermore \hnull\ molecules were reproducibly converted into \up\ by transferring atoms from the Pb tip, which was not possible on \dn\ molecules.
In addition to the differences in constant-current images, the three types of molecules exhibit different conductance spectra as will be detailed below.

In Fig.~\ref{topo}(a), the fraction of \up\ molecules is lower than in (b).
We in fact observed varying fractions of \up, \dn, and \hnull\ molecules in repeated experimental runs, 
presumably because of different substrate temperatures during deposition.

We note in passing that \hnull\ molecules often were isolated rather than embedded in islands.
This observation indicates reduced diffusion of the radical \hnull, which may be due to chemisorption whereas PbPc is expected to be physisorbed like, \eg, Cu phthalocyanine \cite{Kroeger2011}.
Moreover, we verified that \hnull\ molecules can be converted to \hzwei\ by inserting hydrogen from the STM tip as previously demonstrated on Ag(111) \cite{meta_2011}. 

\begin{figure}
\centering \includegraphics[width=0.45\textwidth]{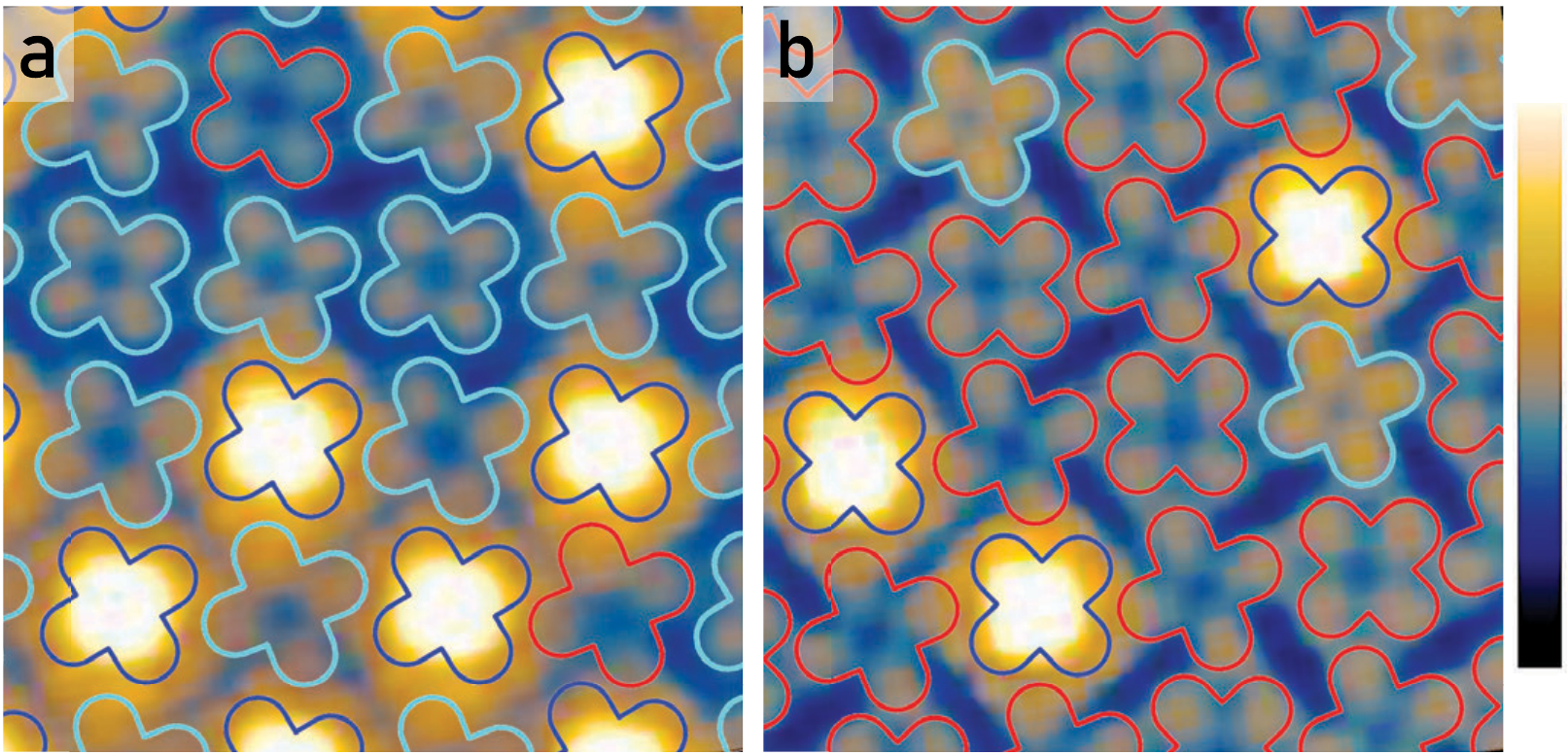}
\caption{
Topographs with molecular outlines indicated.
(a) and (b) show domains A and B\@.
Dark blue, cyan, and red line show \up, \dn, and \hnull\ molecules.
In domain A, the orientations of all molecules are fairly similar while nearest neighbors are rotated by $\approx20^{\circ}$ with respect to each other in domain B\@.
Imaging parameters: $I = 100$\,pA, $V = 6$\,mV in (a) and $-6$\,mV in (b).}
\label{model2}
\end{figure}

\begin{figure}
\centering
\includegraphics[width=0.5\textwidth]{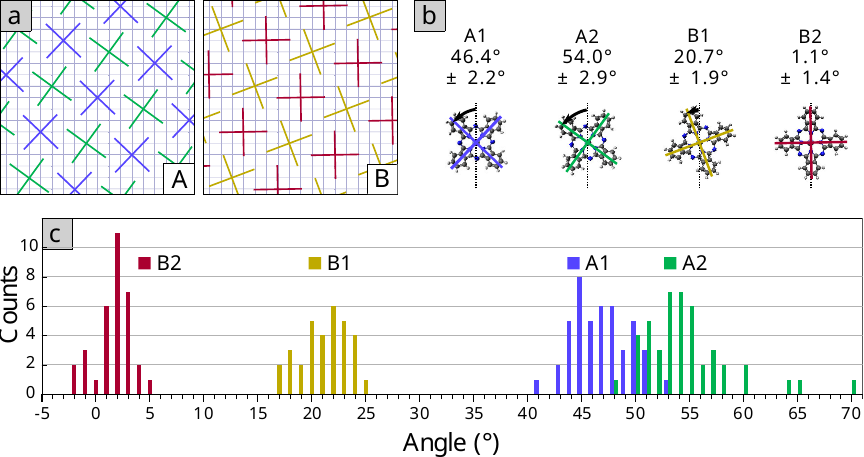}
\caption{Adsorption geometry of PbPc on Pb(100). 
(a) Models of domains A and B\@.
The square Pb(100) mesh is indicated by gray lines.
Phthalocyanine molecules with different orientations are displayed as colored crosses.
The molecules are arranged in a $\begin{bsmallmatrix}5 & 3\\ -3 & 5\end{bsmallmatrix}$ superstructure with two molecules per unit cell and occupy atop positions.
(b) Molecular models of the orientations, which are described by the angles (arrows) enclosed between a molecular lobe and a $\langle110\rangle$ direction.
(c) Histogram of observed angles from 32 B1, 33 B2, 45 A1, and 45 A2 molecules. 
The scatter is partially due to measurement uncertainty ($\approx \pm 2^{\circ}$). 
In few cases (\eg\ $>60^{\circ}$), the molecular orientation does not fall in any of the four classes.}
\label{orientation}
\end{figure}

\paragraph{Structure of Ordered Islands}

In ordered islands, for example Fig.~\ref{topo}(a), we often observed two types of domains.
Close-up views of these domains are displayed in Fig.~\ref{model2} along with outlines indicating the molecular arrangement.
Schematic depictions of the molecular geometries and the substrate mesh are shown in Figure~\ref{orientation}(a). 
We find a checkerboard pattern with two molecules per unit cell, similar to the case of \hzwei\ \cite{Homberg2020}. 
In each domain, the two molecules exhibit different orientations with respect to the underlying Pb(100) lattice leading to a total of four orientations [Figure~\ref{orientation}(b)].
In domain A, we find angles of 46$^{\circ}$(A1) and 54$^{\circ}$ (A2) to a $\langle110\rangle$ direction of the substrate.
In domain B, the angles are 21$^{\circ}$ (B1) and 1$^{\circ}$ (B2).
These values are based on an analysis of more than 125 molecules, which led to the histogram in Figure~\ref{orientation}(c).
The patterns of the domains closely resemble geometries prepared from \hzwei\ on Pb(100) \cite{Homberg2020}.

\section{Electronic Structure}

\subsection{Spectroscopy of Molecular Orbitals}

\begin{figure}[h!]
\centering \includegraphics[width=0.5\textwidth]{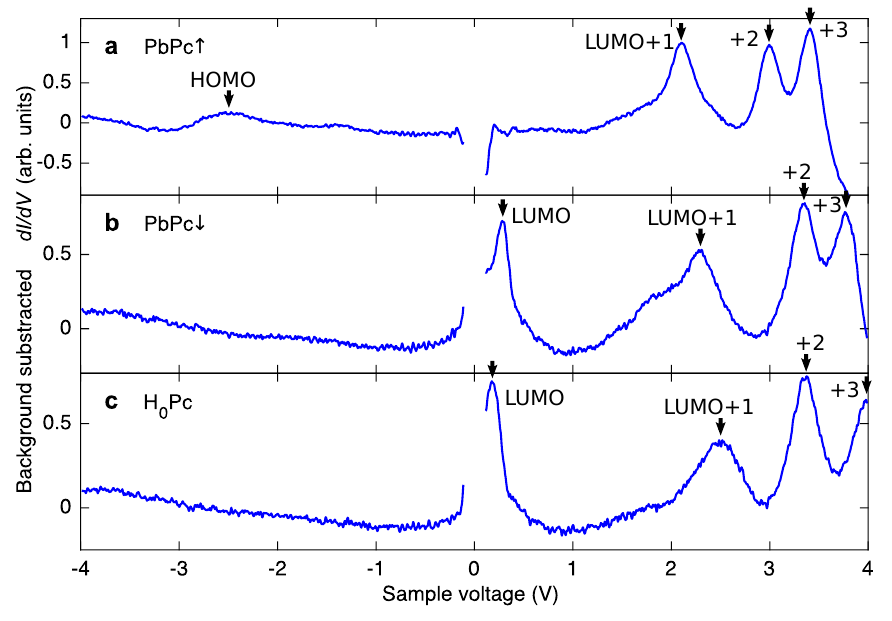}
\caption{\didv\ spectra measured at constant $I = 76$\,pA with the tip placed above (a) \up, (b) \dn, and (c) \hnull\ molecules inside a domain A island.
Data at positive and negative $V$ were measured separately starting at $V = \pm100$\,mV\@.
A voltage modulation $V_M = 14.1$\,mV$_\mathrm{PP}$ was used.
A reference spectrum taken on the Pb(100) substrate has been subtracted from each data set.
Arrows indicate major resonances that we attribute to molecular orbitals.}
\label{lumos}
\end{figure}

Figure~\ref{lumos} shows \didv\ spectra of \up, \dn, and \hnull\ molecules in a type A island.
The measurements were conducted at constant current to enable spectroscopy over a wide voltage range.
This technique also tends to sharpen and slightly down-shift spectral peaks compared to constant-height spectroscopy \cite{Ziegler2009}.
Multiple resonances are observed that we attribute to molecular orbitals. 
The lowest unoccupied molecular orbital (LUMO) of \up \ in islands is very close to $E_F$ and not included in the respective spectrum. 
\dn\ and \hnull\ [Figs.~\ref{lumos} (b) and (c)] exhibit LUMO resonances near 300 and 180\,mV, respectively. 
We use this difference to discriminate between these molecules despite their similar image contrasts.
Three further peaks are observed from all molecules and are attributed to the LUMO+1 to LUMO+3. 
The peak at $V<0$ on \up\ may indicate the HOMO of this molecule.
The weakness of this signal and the absence from the other spectra may be understood from the lateral distribution of the HOMO of phthalocyanines, which vanishes at the molecular center \cite{Baran2010, Lee2010, Jaervinen2014}.
In addition, features at positive sample voltage are usually more intense in tunneling spectroscopy \cite{Griffith1990, Klitsner1990}.

\begin{table}
\begin{tabular}{lrrr}
\hline\hline
Orbital &\up &\dn &\hnull\\
\hline
LUMO+3 &3.4 &3.8 &$\geq4.0$\\
LUMO+2 &3.0 &3.3 &3.4\\
LUMO+1 &2.1 &2.3 &2.5\\
LUMO &$\approx0$ &0.3 &0.2\\
HOMO &$-2.5$\\
\hline\hline
\end{tabular}
\caption{
Energies of the molecular orbitals from spectra of Fig.~\ref{lumos} in eV relative to $E_F$.
The LUMO of \up\ is close to $E_F$ and not discernible in Fig.~\ref{lumos}a.}
\label{lumotabu}
\end{table}

The measured peak energies of the unoccupied orbitals of the three molecules are arranged in fairly similar patterns (Table~\ref{lumotabu}).
The LUMO+1 is separated by $\approx 1$\,eV from the LUMO+2, which in turn lies $\approx 0.5$\,eV below the LUMO+3.
The \up\ orbitals are downshifted by $\approx 0.3$\,eV compared to \dn.
As will be detailed below, this can be qualitatively understood from the molecular electrostatic dipole.
The energy shift is decisive for the appearance of YSR states on \up\ molecules because it leads to a partial occupation of the LUMO\@. 

\begin{figure}[h!]
\centering \includegraphics[width=1.0\columnwidth]{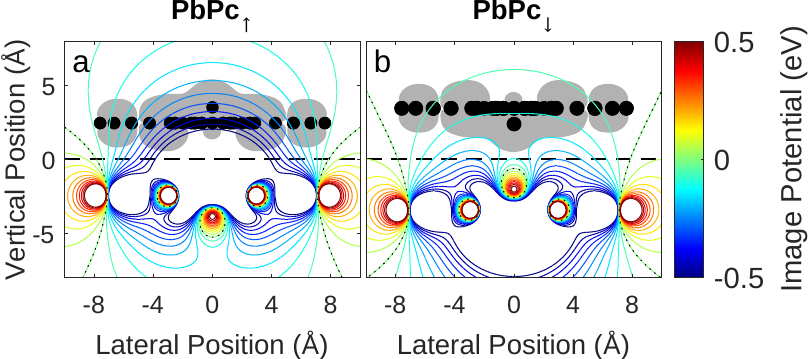}
\caption{Image potential of (a) \up\ and (b) \dn\ molecules calculated from the Mulliken charges from gas-phase DFT calculations.
The image plane used is indicated by a dashed line. 
The molecules are represented by black dots at the projected atomic positions and a gray area representing constant LUMO density.
Colored contour lines range from --500 (blue) to 500\,meV (red) in increments of 50\,meV\@.
$V_{im}=0$ is further marked using black dotted lines.
Negatively charged nitrogen atoms of the molecular macrocycle are accompanied by positive image charges that lower the LUMO energy.
For \dn, the downshift is largely compensated by the negative image charge of the central Pb ion.}
\label{ImagePotential}
\end{figure}

To quantify the electrostatic stray field of PbPc and its influence on orbital energies, gas-phase DFT calculations were performed.
The atoms of the Pc frame were constrained to a plane during structure optimization, while the Pb atom was able to relax in all three directions. 
A Mulliken population analysis reveals negative partial charges of $\approx 0.2$\,e on the nitrogen atoms and finds Pb and macrocyclic carbon atoms to be positively charged by $\approx 0.25$ and 0.13\,e, respectively.
Peripheral C--H bonds are found to be polar with positive (negative) partial charges of $\approx0.08$\,e ($\approx0.06$\,e) on H (C) atoms.
Figure~\ref{ImagePotential} shows the calculated image potential $V_{im}(\mathbf{x})$ of \up\ and \dn\ molecules. 
We used the molecule-substrate distances determined previously from DFT calculations that included the substrate \cite{Homberg2022} and placed the image plane 100\,pm above the topmost substrate nuclei.
The electrostatic shifts of the unoccupied DFT wave functions $\Psi_i$ were evaluated using $\int d^3\mathrm{x} |\Psi_i(\mathbf{x})|^2V_{im}(\mathbf{x})$.
We find a lowering of the energies for \up\ and \dn\ molecules of approximately --300 and --80\,meV, respectively.
The difference between this values matches the experimental shift between the \up\ and \dn\ states (300\,meV) rather well.
The downshifts are caused by the net positive image charge of the C and N atoms of the Pc macrocycle (0.6\,e in total).
The negative image charge of the positive Pb atom partially compensates the downshift, which is the main origin of the different results for \up\ and \dn.

\subsection{Spectroscopy of YSR States}

Below we focus on \up\ molecules in islands because they exhibit spectral features close to the Fermi energy $E_F$ that are consistent with YSR resonances.
Isolated \up\ molecules do not display such features.
Neither do \dn\ or \hnull\ molecules, independent of their surroundings.

\begin{figure}
\centering
\includegraphics[width=0.5\textwidth]{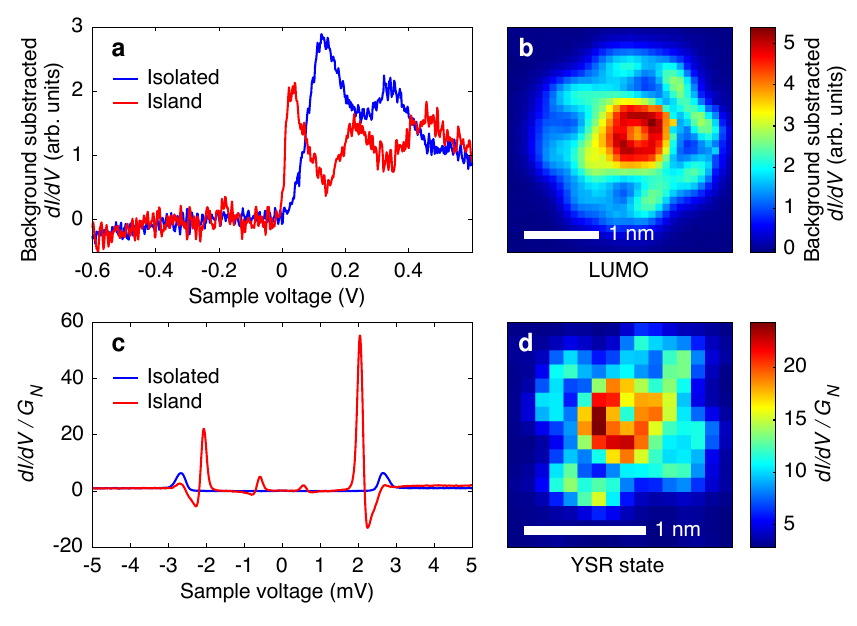}
\caption{(a, c) \didv\ spectra of \up, isolated (blue curves) and inside an ordered island (red curves). 
In (a), a spectrum of the Pb(100) substrate was subtracted for background correction. 
The resonances at low voltages in (a) are attributed to the LUMO\@.
They are located at 140\,mV for the isolated molecule and at 30\,mV for the molecule inside an island.
Additional resonances at higher voltages are likely due to vibronic excitations.
The low-bias spectra in (c) show the coherence peaks of the Pb substrate on the isolated molecule.
In contrast, the spectrum measured inside an island exhibits resonances with different heights at $\approx\pm2.05$\,mV, \ie\ inside the superconductor gap.
The small features at $\approx\pm0.57$\,mV are replica of these peaks due to tunneling of thermally excited electrons or holes.
(b) Spatial map of the LUMO peak height at $V = 135$\,mV of an isolated \up\ molecule.
The map was generated from a $32 \times 32$ grid of $I(V)$ spectra by numerical derivation, background subtraction, and low-pass filtering.
The data are normalized to the normal state conductance $G_N$, which was determined from the constant differential conductance at $V < -3.5$\,mV\@.
(d) Map of the normalized YSR peak height ($V = 2.35$\,mV) of a \up\ molecule inside an ordered island, generated from a $16 \times 16$ grid.
Tip position frozen at (a) $V = 1$\,V, $I = 100$\,pA with a voltage modulation of $V_M = 14.1$\,mV$_\mathrm{PP}$. 
and (c) $V = 6$\,mV, $I = 100$\,pA $V_M = 113\,\mu$V$_\mathrm{PP}$.} 
\label{yesno}
\end{figure}

Figure~\ref{yesno} shows corresponding conductance spectra of \up\ molecules, either isolated (blue) or in an island (red).
The overview spectra in panel (a) reveal that the LUMO energy of \up\ is lowered by $\approx 100$\,meV when the molecule is surrounded by neighbors.
This LUMO shift is accompanied by a drastic change of the low-energy excitations [Figure~\ref{yesno}(c)].
While the spectrum of the isolated molecule (blue) is essentially a copy of the substrate spectrum (not shown)
the molecule with neighbors exhibits YSR peaks near $\pm 2$\,mV\@.
The lateral distributions of the LUMO [Fig.~\ref{yesno}(b)] and the YSR states [Fig.~\ref{yesno}(d)] are similar.
The intensity is highest in a ring shaped area above the Pb ion and some intensity is found on the macrocycle and the lobes. 
This distribution may be understood from the LUMO of PbPc as calculated with density functional theory \cite{Baran2010}.
The LUMO is doubly degenerate and includes hybridized $p_x$ and $p_y$ orbitals of the Pb ion, which in combination give rise to the ring shape. 
The spectroscopic data closely resemble earlier observations from \hzwei\ and indicate that the YSR state is carried by the LUMO\@.

Despite the spectroscopic similarities of \hzwei\ and PbPc, there are also differences.
First, the YSR energies of PbPc cover a wider range, starting from $E_\mathrm{YSR} \approx \Delta$ down to $E_\mathrm{YSR} < \Delta/2$, where $\Delta = 1/2 (\Delta_T + \Delta_S)$ is the average superconductor gap parameter of the tip and the sample.
For \hzwei, the energies were restricted to $E_\mathrm{YSR} \gtrsim 0.9 \Delta$.
Second, the variation of $E_\mathrm{YSR}$ on \hzwei\ was interpreted in terms of the tautomer configuration and the concomitant quadrupole moments of neighboring molecules. 
However, PbPc has no quadrupole moment and consequently the shifts of $E_\mathrm{YSR}$ must have a different origin.

\subsubsection*{Variation of YSR Energies}

\begin{figure}[h!]
\centering \includegraphics[width=0.47\textwidth]{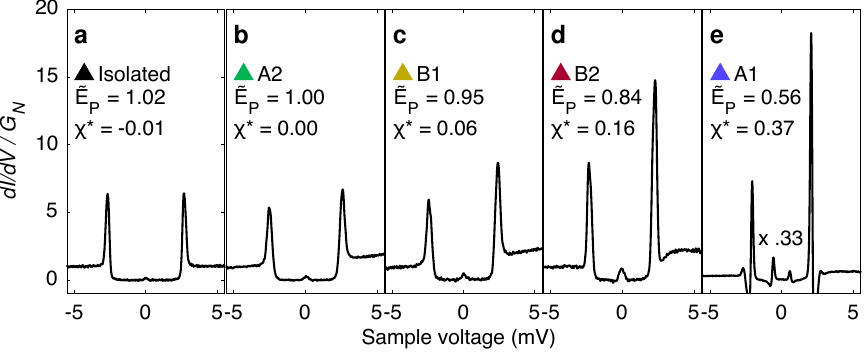}
\caption{\didv\ spectra of \up\ molecules.
(a) No YSR states are observed on isolated \up\ molecules.
(b--e) YSR signatures of molecule classes A1 -- B2. 
The spectrum in (e) is scaled with a factor of 0.33.
All spectra were measured with the tip position frozen at $V = 6$\,mV and $I = 100$\,pA with $V_M = 113\,\mu$V$_\mathrm{PP}.$
The parameters $\tilde{E}_P$ and $\chi^*$ are discussed in the main text.}
\label{smplspx}
\end{figure}

\didv\ spectra we measured for hundreds of \up\ molecules.
Characteristic data are shown in Figure~\ref{smplspx}.
Resolving the YSR and coherence peaks as clearly separated features is only feasible when $E_\mathrm{YSR} \ll \Delta$, \eg, in Figure~\ref{smplspx}(e).
To handle spectra with overlapping peaks, we define a reduced peak energy $\tilde{E}_P = e V_P / \Delta -1$.
Without a YSR state $\tilde{E}_P = 1$, with a pronounced YSR state $\tilde{E}_P \approx \tilde{E}_\mathrm{YSR} = E_\mathrm{YSR} / \Delta$.
A second relevant parameter is the asymmetry $\chi = (P^+ -P^-)/(P^+ +P^-)$ of the peak heights $P^+$ and $P^-$ at positive and negative $V$. 
To determine the intrinsic asymmetry $\chi^*$ a background slope of the spectra was determined at $|eV| \gg \Delta$ and removed by dividing.

\begin{figure}
\centering \includegraphics[width=0.48\textwidth]{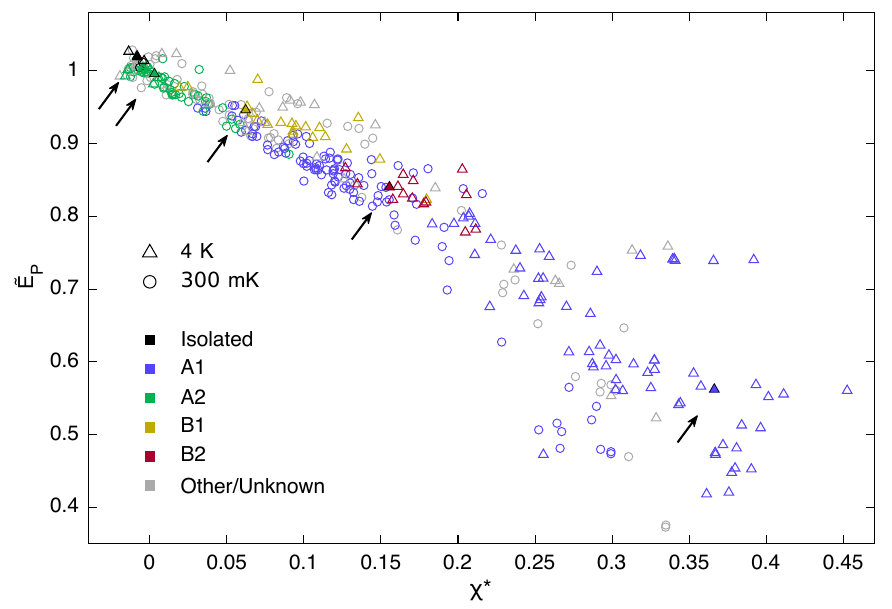}
\caption{
$\tilde{E}_P$ vs.\ $\chi^*$ from many \up\ molecules. 
Data from two STM instruments, denoted 4\,K and 300\,mK, are shown. 
Different molecular configurations are indicated by colors. 
Molecules of groups A1, A2, B1, and B2, were only considered when they were completely surrounded by neighbor molecules.
Data points from molecules at the edges of ordered islands are labeled 'Other'.
The spectra underlying the data points with filled symbols and arrows are shown in Fig.~\ref{smplspx}.}
\label{eofchi}
\end{figure}

An overview of $\tilde{E}_P$ vs.\ $\chi^*$ from many \up\ molecules is presented in Fig.~\ref{eofchi}. 
On the pristine substrate, $\tilde{E}_P = 1$ and $\chi^* = 0$. 
For isolated \up\ molecules (black symbols), we obtain essentially these two values demonstrating the absence of YSR states. 
For \up\ molecules inside ordered islands, however, $\tilde{E}_P$ decreases while $\chi^*$ increases.
In addition, distinct differences are present between the four groups of molecules A1, A2, B1, and B2. 
The strongest YSR states are occur on A1 molecules while A2 molecules only show weak signs of YSR states. 
Although the differences among B1 and B2 molecules are less pronounced the latter molecules display lower YSR energies.

For \hzwei\ enneamers, DFT calculations revealed an electrostatic polarization of the molecular lobes, the polarization being largest within the C-H bonds that point towards neighbor molecules (Figure 9 of Ref.~\citenum{Homberg2020}).
The dipoles in turn electrostatically shift the molecular orbitals.
We suggest that the same effect is at the origin of the different LUMO energies and YSR energies in the case of PbPc.

\subsection{Spatial Mapping of YSR States}

\begin{figure}
\centering \includegraphics[width=0.5\textwidth]{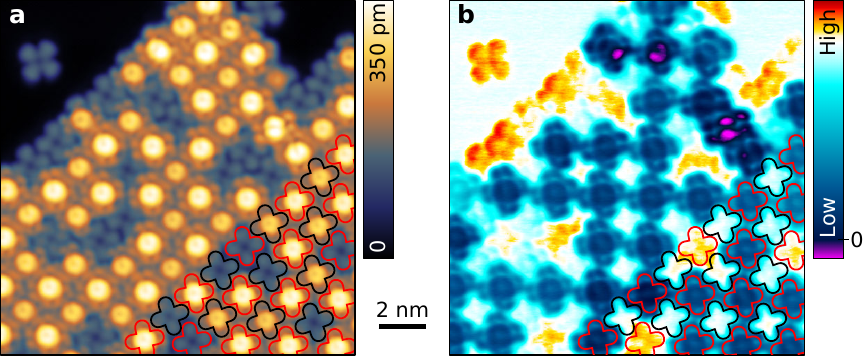}
\caption{(a) Topograph of a domain A island ($V = 2.6$\,mV and $I = 50$\, pA). 
(b) \didv\ map measured simultaneously with the topograph.
Increased (decreased) conductance is shown in yellow and red (blue).
White corresponds to the differential conductance of the substrate.
YSR states reduce the differential conductance at the voltage used and thus appear blue.
Negative differential conductance (purple spots) is occasionally observed. 
The outlines of some molecules (classes A1 and A2, in red and black, respectively) are indicated for easier comparison of the maps.}
\label{didvmap}
\end{figure}

Figure~\ref{didvmap} shows a spatial map of the YSR states.
At the sample voltage $V = 2.6$\,mV, \ie\ just above the coherence peak, YSR states reduce \didv\ (dark colors).
The YSR states are found to be particularly pronounced on group A1 molecules and much weaker on A2.
Other molecules like the isolated \hnull\ in the top left corner exhibit increased differential conductance [yellow and red areas in Fig.~\ref{didvmap}(b)]. 
This effect may be due to the LUMO energy of the molecule, which is higher than for molecules inside the island.
However, we hint that another factor may play a role.
We observed that the molecules lacking a YSR state exhibit a slightly larger superconducting gap.
For example, isolated \up\ molecules, showed values of up to $\tilde{E}_p \approx 1.06$ (Fig.~\ref{eofchi}). 
An apparent change of the superconducting gap may actually result from the two-band superconductivity of Pb \cite{ruby_experimental_2015}. 
An increased gap size shifts the coherence peak closer to the sample voltage used and thus increases the conductance.

The spatial distribution of the YSR features appears to be twofold symmetric on some of the molecules. 
This reduction from the fourfold symmetry of PbPc may be due to a reduced symmetry of the environment, which lifts the degeneracy of the PbPc LUMO\@. 
The YSR state is then expected to be carried by the lower of the twofold symmetric orbitals.

The presence or absence of Pb ions introduces a degree of randomness in molecular islands, despite the regular lateral arrangement of the molecules. 
Similarly, Figure~\ref{eofchi} reveals variations of the YSR state energies among different groups of molecules, but also within a single group.
For example, the YSR energies in group A1 scatter between 0.4 $\Delta$ and 0.95 $\Delta$. 
Below we relate this variability to differences in the shell of neighbors.

In ordered islands consisting of \up, \dn, and \hnull\ molecules, the shells of nearest and next nearest neighbors (NN and NNN, respectively) may contain any combination of these constituents.
We define a model energy
\begin{equation}
\tilde{E}_{P,model} = \tilde{E}_{P,0} + \sum_n \epsilon(t_n,d_n).
\label{para}
\end{equation}
$\tilde{E}_{P,0}$ is a starting value.
The environment introduces shifts $\epsilon(t_n,d_n)$ that depend on the type $t_n$ (\up, \dn, or \hnull) of a neighbor and its distance $d_n$ (NN or NNN).
The summation includes NN and NNN molecules.

\begin{figure}[h!]
\centering \includegraphics[width=0.4\textwidth]{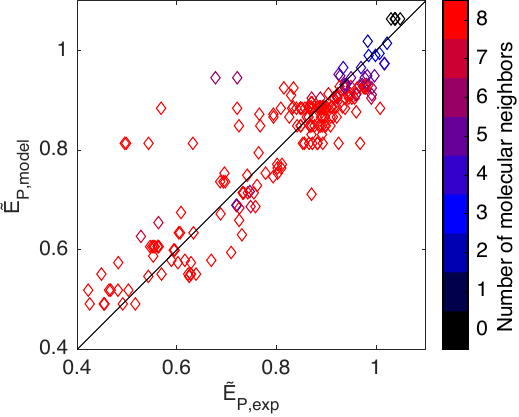}
\caption{Parametrization of the peak energy.
Data from A1 \up\ molecules and isolated \up\ molecules are shown. 
The black line indicates a fit according to Eq.~\ref{para}.}
\label{neigh}
\end{figure}

\begin{table}
\begin{tabular}{lrr}
\hline\hline
Type 			&NN 								&NNN\\
\hline
\up				&$-25 \pm 22$ 	&$-20 \pm 16$\\
\dn				&$-102 \pm 24$ 	&$-48 \pm 22$\\
\hnull		&$-42 \pm 24$ 	&$-00 \pm 18$\\
\hline\hline
\end{tabular}
\caption{Fit of model parameters $\epsilon(t, d)$ in \textperthousand. 
For an isolated molecule, the fit leads to $E_{p,0} = 1.064 \pm 0.057$. 
Margins represent 95\% confidence intervals from the fit.
NN and NNN \dn\ molecules have a significant impact on the YSR energy of a \up\ molecule.}
\label{tabu}
\end{table}

Table~\ref{tabu} shows the results of a fit (Fig.~\ref{neigh}) of this model to the experimental data.
As expected, each additional molecular neighbor reduces the peak energy. 
\dn\ neighbors have a large impact on the peak energy $\tilde{E}_P$ whereas the the influence of \up\ and \hnull\ neighbors is less significant.
States with the lowest $\tilde{E}_P$ of $\approx 0.4$ are observed when all NN and NNN sites are occupied by \dn\ molecules. 

At first glance, an anti-parallel alignment of the electrostatic dipole of the central molecule with respect to the neighboring dipoles might be expected to lower its LUMO energy. 
However, a calculation of the effects of the electrostatic stray fields of \up\ and \dn\ neighbors predicts only minor differences because the main contribution to the field is due the positive hydrogen atoms at the periphery.
They electrostatically shift the LUMO of NN molecules by 33\,meV when image charges are neglected and slightly less when image charges are considered (20 and 29\,meV for \dn\ and \up, respectively). 
These values are similar to the electrostatic shifts previously found for H$_2$Pc~\cite{Homberg2020}.
We speculate that the different shifts caused by \up\ and \dn\ neighbors may be be related to their different LUMO energies.
The \up\ LUMO being closer to $E_F$, a partial occupation appears likely.
The corresponding negative charge is expected to counteract the effect of the positive hydrogen ions.
A quantitative estimate would required additional data on the molecular geometries in the adsorbed states and the position of the image plane.
Moreover, a microscopic description of screening may be necessary.\\

\section{Summary}

Similar to \hzwei, PbPc is diamagnetic in gas-phase but becomes paramagnetic upon assembly into suitable clusters on a Pb(100) surface. 
This is reflected by pronounced YSR resonances inside the superconductor gap of Pb, typically at lower energies than for \hzwei. 
The paramagnetism results from electrostatic shifts of the LUMO that are caused by horizontal induced dipole moments in a molecule and its interacting neighbors.
In particular, arrays that are disordered with respect to the up and down orientation of the PbPc molecules display a range of YSR state energies.
A statistical analysis is used to quantify the influence of nearest and next-nearest neighbor molecules.
The interaction of the charge distribution of PbPc molecules with its image charge leads to a shift between the orbitals of \up\ and \dn\ molecules.

As of today, three phthalocyanines were found to exhibit electrostatically induced paramagnetism on a substrate.
Their common features are a LUMO that is fairly close to the Fermi level and a heterocycle that provides partial charges for intermolecular electrostatic interaction.
These requirements are probably met by a range of closed-shell molecules that likely may be made paramagnetic on substrates.

%

\end{document}